\documentclass[osajnl,twocolumn,showpacs,superscriptaddress,10pt]{revtex4} 
\usepackage{amsmath,amssymb,graphicx}
\begin{document}

\title{Octonacci Photonic Quasicrystals}

\author{E.R. Brand\~ao}
\author{C.H. Costa}\email{Corresponding author: carloshocosta@hotmail.com}
\author{M.S. Vasconcelos}\email{Corresponding author: mvasconcelos@ect.ufrn.br}
\affiliation{Escola de Ci\^encia e Tecnologia, Universidade Federal
do Rio Grande do Norte, 59072-970, Natal-RN, Brazil}
\author{D.H.A.L. Anselmo}
\affiliation{Departamento de F\'{\i}sica Te\'orica e Experimental,
Universidade Federal do Rio Grande do Norte, Natal-RN, 59072-970,
Brazil}
\author{V.D. Mello}
\affiliation{Departamento de F\'isica, Universidade do Estado do Rio Grande do Norte, 59610-210, Mossor\'o-RN, Brazil}

\begin{abstract}
We study theoretically the transmission spectra in one-dimensional photonic quasicrystals, made up of SiO$_2$($A$) and TiO$_2$($B$) materials, organized following the Octonacci sequence, where the $n$th-stage of the multilayer $S_{n}$ is given by the rule $S_{n}=S_{n-1}S_{n-2}S_{n-1}$, for $n\geq 3$ and with $S_{1}=A$ and $S_{2}=B$. The expression for transmittance was obtained by employing a theoretical calculation based in the transfer-matrix method. To normally incident waves, we observe that, for a same generation, the transmission spectra for TE and TM waves are equal, at least qualitatively, and they present a scaling property where a self-similar behavior is obtained, as an evidence that these spectra are fractals. The spectra show regions where the omnidirectional band gaps emerges for specific generations of Octonacci photonic structure, except to TM waves. For TE waves, we note that all of them have the almost same width, for different generations. We also report the localization of modes as a consequence of the quasiperiodicity of the heterostructure.
\end{abstract}


\maketitle 

Since the pioneer works of E. Yablonovitch\cite{Yablonovitch} and S. John\cite{John} in which the photonic crystals (PCs) were proposed, many experimental and theoretical investigations have been devoted to the understanding of the physical properties of these systems (for a review, we recommend the Refs.\cite{Prather,Laine,Massaro}). PCs are structures characterized by the periodic variation of their refractive index, allowing the appearance of the so-called photonic band gaps (PBGs)\cite{Sakoda,Joannopoulos}, and, because of this, they have the property to control the light propagation, leading to a new age for optical devices\cite{Cerqueira,Macia}. Some very interesting applications of PBGs are in the fields of waveguides
based in PC\cite{Kramper}, optical reflectors\cite{Chigrin}, etc.

On the other hand, the discovery of quasicrystals, in 1982, by Shechtman \textit{et al.}\cite{Shechtman} has fired up a new field of condensed matter physics, resulting in many theoretical and experimental works (for an up to date review of this field, see Refs.\cite{Eudenilson2,Macia}). For such amazing discovery, in 2011, Shechtman was awarded with the Nobel Prize in Chemistry\cite{Shechtman2}. Historically, the concept of quasiperiodicity was first introduced by the mathematician H. A. Bohr, in 1926, during his studies about almost periodic functions\cite{Bohr}. Decades later, in 1970, the works of Penrose and Ammann demonstrated that it is possible to fill a non-Euclidean gapless space in a nonperiodic way. These analysis can be considered as the foundations of the field of quasiperiodic structures (or tillings)\cite{Steurer}. Penrose has proposed a tiling of the plane,
which today is well known as the \emph{Penrose tiling}, with two rhombic tiles which are arranged according to specific
rules\cite{Penrose}. In the same decade, Ammann found tilings with
an eight-fold symmetry filling the gapless
plane\cite{Grunbaum,Ammann}. Today, many different tilings are known
which can not only fill the two-dimensional space quasiperiodically,
but also three and higher dimensions, like the three-dimensional
Ammann-Kramer-Neri tiling\cite{Kramer}  or the
zonotiles\cite{Towle}.

For a quasiperiodic arrangement, which displays a important role for
promising technological applications of aperiodic
systems\cite{Macia2}, the photonic crystals can be called photonic
quasicrystals (PQCs)\cite{Steurer2,Vardeny}, which are defined like
the PCs: the difference is in the spatial distribution of the
refraction indexes, where they are organized in a quasiperiodic
fashion\cite{Costa}, obeying a mathematical rule, which, here, is
the Octonacci substitutional sequence. The studies of photonic
quasicrystals started with the pioneer works of Kohomoto \textit{et
al.}\cite{Kohmoto}, where they proposed a one-dimensional
multilayered system arranged in accordance to a Fibonacci
quasiperiodic rule, with a quarter wavelength condition. Since that,
many theoretical and experimental works have been published in this
field (see Refs.\cite{Zentgraf,Utikal} and references therein). It
has been recognized that in addition to crystalline and amorphous
materials, there exists a third intermediate class known as
``deterministic aperiodic" structures, which can be generated by a
substitution rule based on two or
more\cite{Barber_book_2008,dal_Negro_book_2013} building blocks that
exhibit long-range order, which does not have translational
symmetry. Also, it has been shown that these structures exhibit
properties of self-similarity in their spectra\cite{Eudenilson2}. In
a general way, this new class of structures can be classified into
two groups: quasicrystals and all other deterministic aperiodic
structures. Quasicrystals therefore represent a special class of
deterministic aperiodic structures. A more precise and updated
definition of quasicrystals with dimensionality $n$ ($n$ = 1, 2 or
3) is that in addition to their possible generation by a
substitution process, they can also be formed from a projection of
an appropriate periodic structure in a higher dimensional space
$mD$, where $m > n$\cite{Vardeny}. In contrast, structures that are
part of the other deterministic structures cannot be constructed in
such a manner. For example, in one dimension (1D), quasicrystalline
structures include the Fibonacci sequence and their
generalizations\cite{Costa3,Costa4}. Examples of aperiodic
structures that differ from quasicrystals are systems that obey the
Thue-Morse, double-period\cite{Vasconcelos} and
Rudin-Shapiro\cite{Vasconcelos3} sequences.

The aim of this work is to investigate the transmission spectra of a
obliquely incident light beam from a transparent medium into a
multilayer photonic structure composed of SiO$_{2}$ and TiO$_2$
layers arranged in accordance to the quasiperiodic Octonacci
sequence, which describes the arrangement of spacing of the Ammann
quasilattice (8-grid), namely, the octagonal Ammann-Becker
tiling\cite{Steurer}.

This paper is organized as follow: in Sec. \ref{sec:Theoretical
Method}, we present the Octonacci sequence, as well the theoretical
model to calculate the transmission spectra of TE and TM waves in
PQCs, which is based in well-known transfer-matrix method. In Sec.
\ref{sec:Numerical Results}, we present some numerical results for
the transmittance spectra as function of frequency and incident
angle of the wave. We could observe the expected self-similar
behavior of the transmission spectra for normally incident waves and
the appearance of the omnidirectional photonic band gaps, that are
well defined only for TE waves. The conclusions are presented in
Sec. \ref{sec:Conclusions}. Finally, we present our
acknowledgements.

\section{Theoretical Model: Octonacci Sequence and Transfer-Matrix Method}\label{sec:Theoretical Method}

An Octonacci multilayer photonic structure can be defined by by the
growth, by juxtaposition, of two building blocks
\textit{A}(SiO$_{2}$) and \textit{B}(TiO$_{2}$), where the
$n$th-stage of the multilayer $S_{n}$ is given iteratively by the
rule\cite{prb62}
\begin{equation}\label{eq:Sn}
S_{n}=S_{n-1}S_{n-2}S_{n-1},
\end{equation}
for $n\geq 3$, with $S_{1}=A$ and $S_{2}=B$. The number of the
building blocks increases according to Pell number
$P_{n}=2P_{n-1}+P_{n-2}$ (with $P_{1}=1$, $P_{2}=1$ and $n\geq3$).
The number of the building blocks of the materials $A$ and $B$ in
$n$-th stage is given by $n_{A}+n_{B}=P_{n}$. Here, $n_A= P_n$ and
$n_B = P_{n-1}$, with $P_{1}=0$, $P_{2}=1$ and $n\geq3$. The number
of building blocks $B$ divided by the number of building blocks $A$,
in the limit $n\rightarrow \infty$, is $\tau=1+\sqrt{2}$. Another
way to obtain this sequence is by using the follow recurrence rule:
\textit{A}$\rightarrow$\textit{B},
\textit{B}$\rightarrow$\textit{BAB}. Note that this sequence is
classified as Pisot-Vijayaraghavan (PV), when we take the negative
eigenvalue of the substitution matrix\cite{Lu}, i e., $\sigma^{-}=
1- \sqrt{2}$, with $|\sigma^{-}| \le 1$. On other hand, differently
of the Fibonacci case,  the ratio $\tau$ between the number of
building blocks $A$ and building blocks $B$ is different of the
positive eigenvalue $\sigma^{+}=1+ \sqrt{2}$, as usually is found in
the Fibonacci generalizations\cite{Costa,Barriuso}.

Although several theoretical techniques have been used to study the
transmission spectra in these structures, like spectral
analysis\cite{Golmohammadi}, in the present work we make use of the
transfer-matrix approach to analyze them (for a review see
Ref.\cite{Eudenilson2}). On the present section we have two goals:
first, we want to investigate the behavior of the normally incident
light when it pass through a Octonacci photonic layered system,
considering the central wavelength $\lambda_{0}=700$ nm, where the
central gap will appear\cite{Vasconcelos}. Second, we intend to
investigate the influence of the oblique incidence in such systems,
by looking for frequency regions where the band gaps are independent
of the incident angle and/or of the wave polarization, i.e., the
so-called omnidirectional photonic band gaps.

\begin{figure}[h]
\centering
\includegraphics[width=\columnwidth]{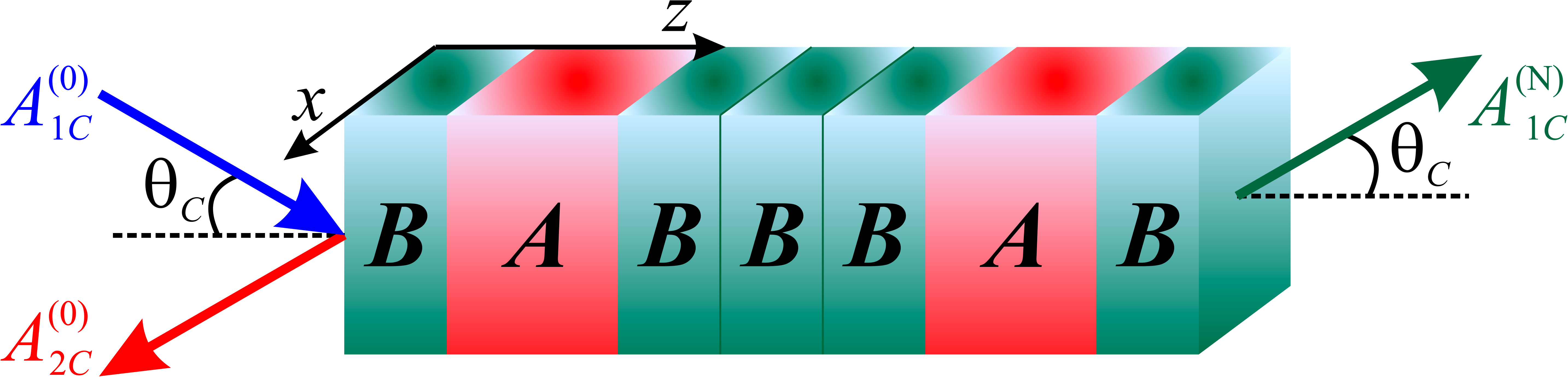}
\caption{Schematic representation for the geometry of the Octonacci
quasiperiodic multilayer system considered in this work, more
precisely, for sequence $S_{4}=[B|A|B|B|B|A|B]$, with $P_{4}=7$. $L$
is the size of the whole superlattice structure.}\label{fig1}
\end{figure}

We consider that an electromagnetic radiation of frequency $\omega$
and {\it s}-polarization (TE waves), or {\it p}-polarization (TM
waves), is incident from a transparent medium $C$ at an arbitrary
angle $\theta_{C}$ with respect to the normal direction of the
layered system (see Fig. \ref{fig1}). The layered system is formed
by an array of slabs of the different materials $A$ and $B$. The
reflectance $R$ and the transmittance $T$ coefficients are given,
respectively, by\cite{Vasconcelos3},
\begin{equation}\label{eq:r_t}
R=\left|\frac{M_{n}(2,1)}{M_{n}(1,1)}\right|^{2} \qquad \textrm{and} \qquad T=\left|\frac{1}{M_{n}(1,1)}\right|^{2},
\end{equation}
where $M_{n}(i,j)$, with $i$,$j=1,2$, are the elements of the
optical transfer-matrix $M_{n}$, for the $n$-th generation of the
sequence, and which links the coefficients of the electromagnetic
fields in the region $z < 0$ to those in the region $z > L$, where
$L$ means the size of the quasiperiodic structure.

In order to illustrate the method to calculate the optical
transfer-matrices, let us consider, firstly, the case in which we
have two different dielectric media $A$ and $B$ with thicknesses
$d_{A}$ and $d_{B}$, and non-dispersive refractive indexes $n_{A}$
and $n_{B}$, respectively, and a third transparent medium $C$, with
refractive index $n_{c}$ that surrounds the quasiperiodic unit cell
composed by the binary chain $A$ and $B$ (see Fig. \ref{fig1}). The
transmission of an obliquely incident light wave across the
interfaces $\alpha\rightarrow\beta$ (i.e., $C\rightarrow A$,
$A\rightarrow B$, $\cdots$, $B\rightarrow C$) is represented by the
transmission matrix (for more details, see Ref.\cite{Zhan})
\begin{equation}\label{eq:ms}
\displaystyle M^{s}_{\alpha\rightarrow \beta}=\frac{1}{2}\left[\begin{array}{rr}
1+\frac{k_{z\beta}}{k_{z\alpha}}&1-\frac{k_{z\beta}}{k_{z\alpha}}\\
1-\frac{k_{z\beta}}{k_{z\alpha}}&1+\frac{k_{z\beta}}{k_{z\alpha}}
\end{array}\right],
\end{equation}
with
\begin{equation}\label{eq:kz}
\displaystyle k_{z\alpha}=\left[\left(\frac{n_{\alpha}\omega}{c}\right)^{2}-k_{x}^{2}\right]^{1/2},
\end{equation}
and
\begin{equation}\label{eq:kx}
\displaystyle k_{x}=n_{c}\frac{\omega}{c}\sin(\theta_{C}),
\end{equation}
for TE or $s$-polarization waves. Beside of this, for TM or $p$-polarization waves crossing the interface $\alpha|\beta$, from medium $\alpha$ to $\beta$ one, we have
\begin{equation}\label{eq:mp}
\displaystyle M^{p}_{\alpha\rightarrow \beta}=\frac{1}{2}\frac{n_{\beta}}{n_{\alpha}}\left[\begin{array}{rr}
1+\frac{k_{z\beta}}{k_{z\alpha}}\frac{n_{\alpha}}{n_{\beta}}&1-\frac{k_{z\beta}}{k_{z\alpha}}\frac{n_{\alpha}}{n_{\beta}}\\
1-\frac{k_{z\beta}}{k_{z\alpha}}\frac{n_{\alpha}}{n_{\beta}}&1+\frac{k_{z\beta}}{k_{z\alpha}}\frac{n_{\alpha}}{n_{\beta}}
\end{array}\right].
\end{equation}

On the other hand, the propagation of the light wave within of a layer $\gamma$ ($\gamma=A$ or $B$), for both TE and TM waves, is characterized by the propagation matrix\cite{Zhan}
\begin{equation}\label{eq:mg}
\displaystyle M_{\gamma}=\left[\begin{array}{rr}
 \exp(-ik_{z,\gamma}d_{\gamma}) &0 \\
 0 & \exp(ik_{z,\gamma}d_{\gamma})
\end{array}\right],
\end{equation}
where $d_{\gamma}$ is the thickness of the respective material.

For TE waves, we assume that, for a given generation of the
Octonacci sequence, the electrical field for $j$-th slab (where
$j=0,1,2,\cdots,N$, and $N$ is the total number of slabs in the unit
cell) has the form
$\vec{\textbf{E}}_{\gamma}^{(j)}=\left(0,E_{y,\gamma}^{(j)},0\right)$
with
\begin{equation}
E_{y,\gamma}^{(j)}=\left[A_{\gamma}^{(j)}\exp(ik_{z,\gamma}z)+B_{\gamma}^{(j)}\exp(ik_{z,\gamma}z)\right]\exp{i(k_{x}x-\omega t)},
\end{equation}
where $A_{\gamma}^{(j)}$ and $B_{\gamma}^{(j)}$ ($\gamma=A$ or $B$)
are the field amplitudes in the $j$-th slab. On the other hand, for
the case of TM waves, we assume that the magnetic field for $j$-th
slab os given by
$\vec{\textbf{H}}_{\gamma}^{(j)}=\left(0,H_{y,\gamma}^{(j)},0\right)$
with
\begin{equation}
H_{y,\gamma}^{(j)}=n_{\gamma}\left[A_{\gamma}^{(j)}\exp(ik_{z,\gamma}z)+B_{\gamma}^{(j)}\exp(ik_{z,\gamma}z)\right]\exp{i(k_{x}x-\omega t)}.
\end{equation}

By applying the Maxwell's electromagnetic boundary conditions at the
interfaces, to successive multiplications of the transmission
$M_{\alpha\rightarrow\beta}$ and propagation $M_{\gamma}$ matrices
along the finite structure, we obtain
\begin{equation}
 \left(\begin{array}{c}
    A_{1C}^{(0)}\\A_{2C}^{(0)}
\end{array}\right)=M_{n}\left(\begin{array}{c}
    A_{1C}^{(N)}\\0
\end{array}\right),
\end{equation}
where the transfer-matrix for the $n$-th generation $M_{n}$ is a $2\times2$ matrix given by
\begin{equation}\label{eq:transfer matriz}
M_{n}=M_{CB}M_{B}M_{BA}M_{A}M_{AB}M_{B}\cdots M_{B}M_{BC}.
\end{equation}

Now, by considering the iterative rule given in Eq. (\ref{eq:Sn}), we are capable to find the following general formula for the transfer-matrices in the Octonacci quasiperiodic system,
\begin{equation}
M_{n}=M_{CB}T_{n}M_{BC},\qquad \textrm{(for $n\geq 3$)}
\end{equation}
where
\begin{equation}
T_{n}=T_{n-1}T_{n-2}T_{n-1},
\end{equation}
whose initial conditions are $M_{1}=T_{CB}T_{1}T_{BC}$, $M_{2}=T_{CB}T_{2}T_{BC}$, $T_{1}=M_{B}$, and\linebreak $T_{2}=M_{B}M_{BA}M_{A}M_{AB}M_{B}$.

\section{Numerical Results}\label{sec:Numerical Results}

Now we present our numerical results to illustrate the optical transmission spectra of some quasiperiodic structures. We consider the same physical parameters used in Ref.\cite{Vasconcelos}, i.e., those appropriated for silicon dioxide ($A$) and titanium dioxide ($B$); they are virtually absorption-free above 400 nm\cite{Gellermann}. Also, we consider the individual layers as quarter-wave layers, for which the quasiperiodicity is expected to be more effective\cite{Vasconcelos3}, with the central wavelength $\lambda_{0}=700$ nm. These conditions yield the physical thickness $d_{\gamma}=(700/4 n_\gamma)$ nm, $\gamma=A$ or $B$, such that $n_{A}d_{A}=n_{B}d_{B}$. Their dielectric constants around the central wavelength $\lambda_{0}=700$ nm are $n_{A} = 1.45$ and $n_{B} = 2.30$, respectively. We also consider medium $C$ to be vacuum, and the phase shifts are given by:
\begin{equation}
\delta_{A}= (\pi/2) \Omega \cos(\theta_{A});\,\,\,\,\,\
\delta_{B}=(\pi/2) \Omega \cos(\theta_{B})
\end{equation}
\noindent where $\Omega$ is the reduced frequency  $\omega /\omega_0= \lambda_{0}/\lambda$\cite{Vasconcelos}.

\begin{figure}[h]
\centering
\includegraphics[width=\columnwidth]{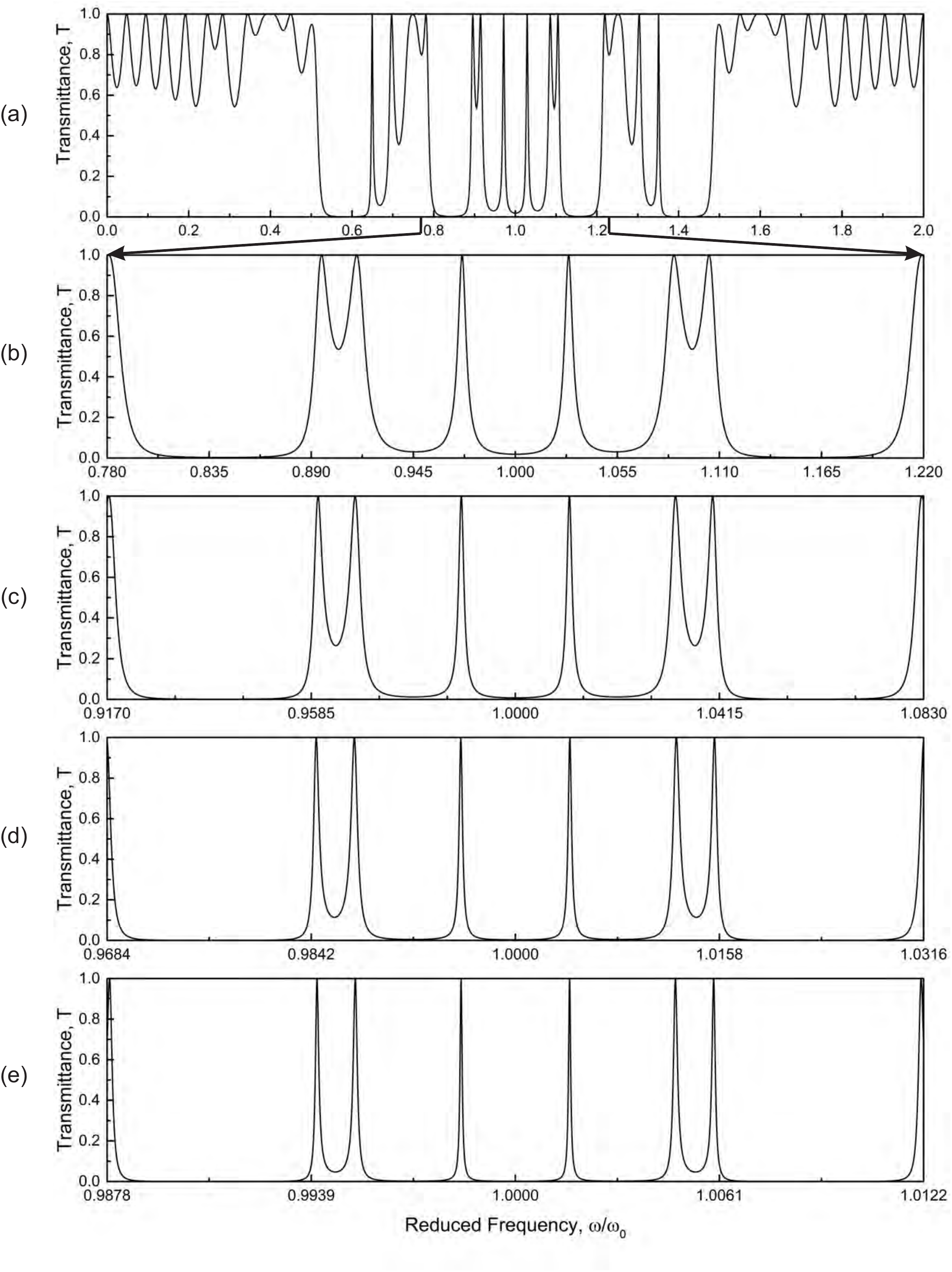}
\caption{Transmission spectra as a function of the reduced frequency $\omega/\omega_{0}$ for normal incidence of TE waves: (a) fifth generation of Octonacci sequence ($P_{5}=17$ layers); (b) same as (a), but for the reduced range of frequency $0.78\le \omega/\omega_{0}\le 1.22$; in (c), (d) and (e) we have the same as (a), but for sixth ($P_{6}=41$ layers), seventh ($P_{7}=99$ layers) and eighth ($P_{8}=239$ layers) generations of Octonacci sequence, with the respective scales $0.917\le \omega/\omega_{0}\le 1.083$, $0.9684\le \omega/\omega_{0}\le 1.0316$, and $0.9878\le \omega/\omega_{0}\le 1.0122$.}\label{fig2}
\end{figure}

The light transmission spectrum, as a function of reduced frequency $\omega/\omega_{0}$, for the fifth generation of the quasiperiodic Octonacci sequence (where 
$P_{5}=17$ layers) is presented in Figs. \ref{fig2}a, for TE waves and a normally incident wave, i.e., $\theta_{C}=90^{\circ}$. As it is expected for normal incidence, the transmission spectrum has a mirror symmetry around the mid-gap reduced frequency $\omega/\omega_{0}=1$ because the quarter-wavelength condition for a periodic multilayer was taken in account. However, differently of what Vasconcelos and Albuquerque reported about a Fibonacci photonic quasicrystal\cite{Vasconcelos}, the light is almost entirely reflected and the structure is quite opaque (the transmission coefficient is less than 0.02 for both wave polarizations) at that frequency. This happens because, from a wave point of view, the layers $A$ and $B$ are not equivalent. The transmission spectrum for TM waves is not displayed here because it is qualitatively and quantitatively equal to TE case.

Furthermore, as usually occur for any wave phenomena in quasiperiodic heterostructures, the transmission spectrum presents a scaling property with respect to the generation number of Octonacci sequence, within a symmetrical interval around the mid-gap reduced frequency $\omega/\omega_{0}=1$. To make this more clear, we present in Fig. \ref{fig2}b the transmission spectra shown in Figs. \ref{fig2}a, but for the reduced range of frequency. This spectrum is similar to the one representing the sixth generation of the quasiperiodic Octonacci sequence (where $P_{6}=41$ layers), displayed in Fig. \ref{fig2}c for the range of frequency reduced by a scale factor approximately equal to 2.6. In Figs. \ref{fig2}d and \ref{fig2}e, we present the transmission spectra for seventh and eighth generations (where $P_{7}=99$ and $P_{8}=239$ layers), and we can observe that such spectra repeat every generation for a Octonacci unidimensional photonic quasicrystal, while in the Fibonacci sequence case, the spectra repeat each six generations, when the optical parameters of medium $C$ are different from $A$\cite{Vasconcelos}, and each three generations, when we take medium $C$ equal to $A$ one\cite{Kohmoto}. This scaling property corresponds to a self-similar behavior of the spectrum, which is a qualitative evidence of a fractal spectrum.

\begin{figure}[h]
\centering
\includegraphics[width=\columnwidth]{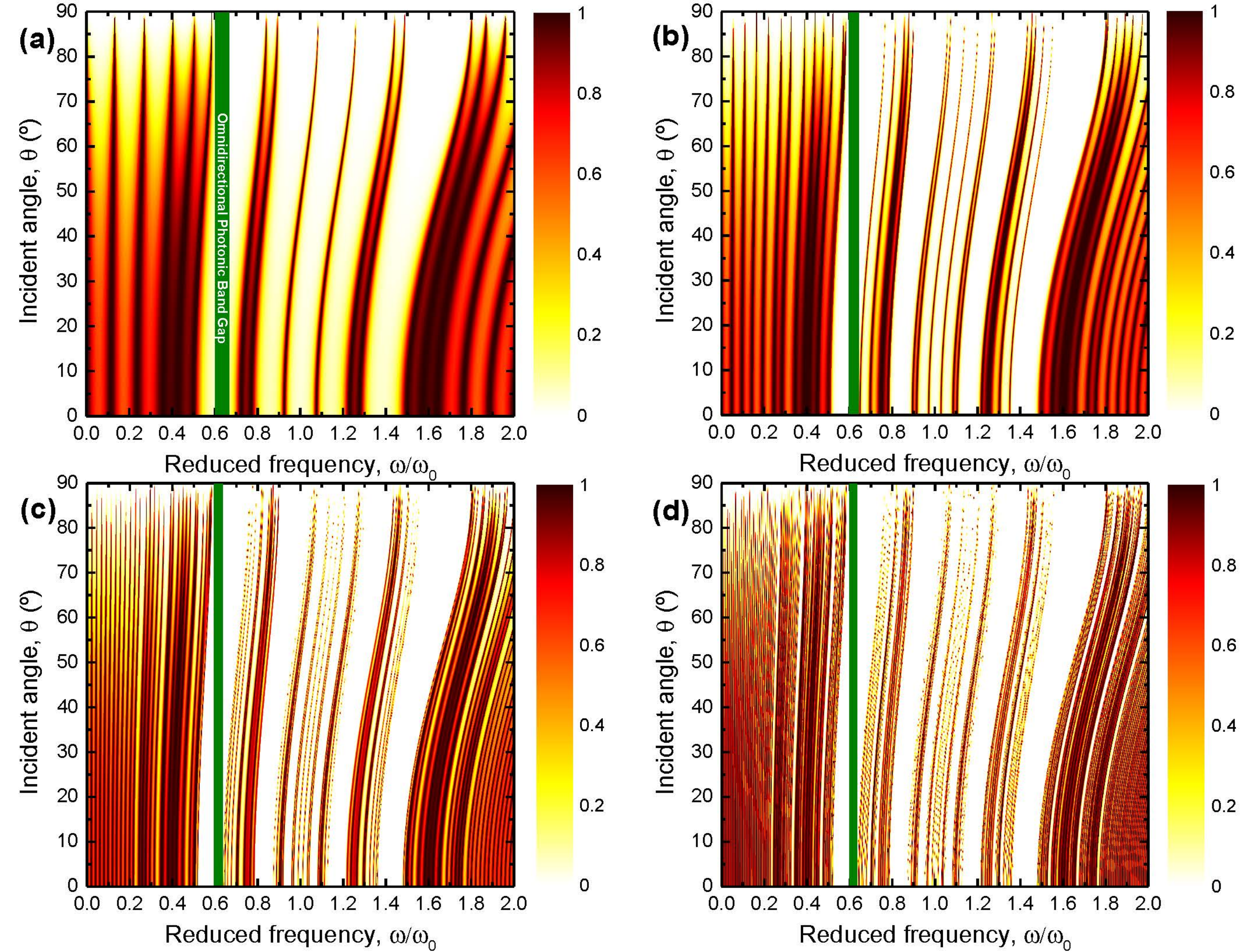}
\caption{Transmission spectra as a function of reduced frequency $\omega/\omega_{0}$ and the incident angle $\theta_{C}$ for TE waves for: (a) fourth; (b) fifth; (c) sixth; and (d) seventh generations of Octonacci sequence. In all these figures the white (black) color means a transmission coefficient equal to 0 (1). The omnidirectional photonic band gap is enhanced by a green rectangle.}\label{fig3}
\end{figure}

\begin{figure}[h]
\centering
\includegraphics[width=\columnwidth]{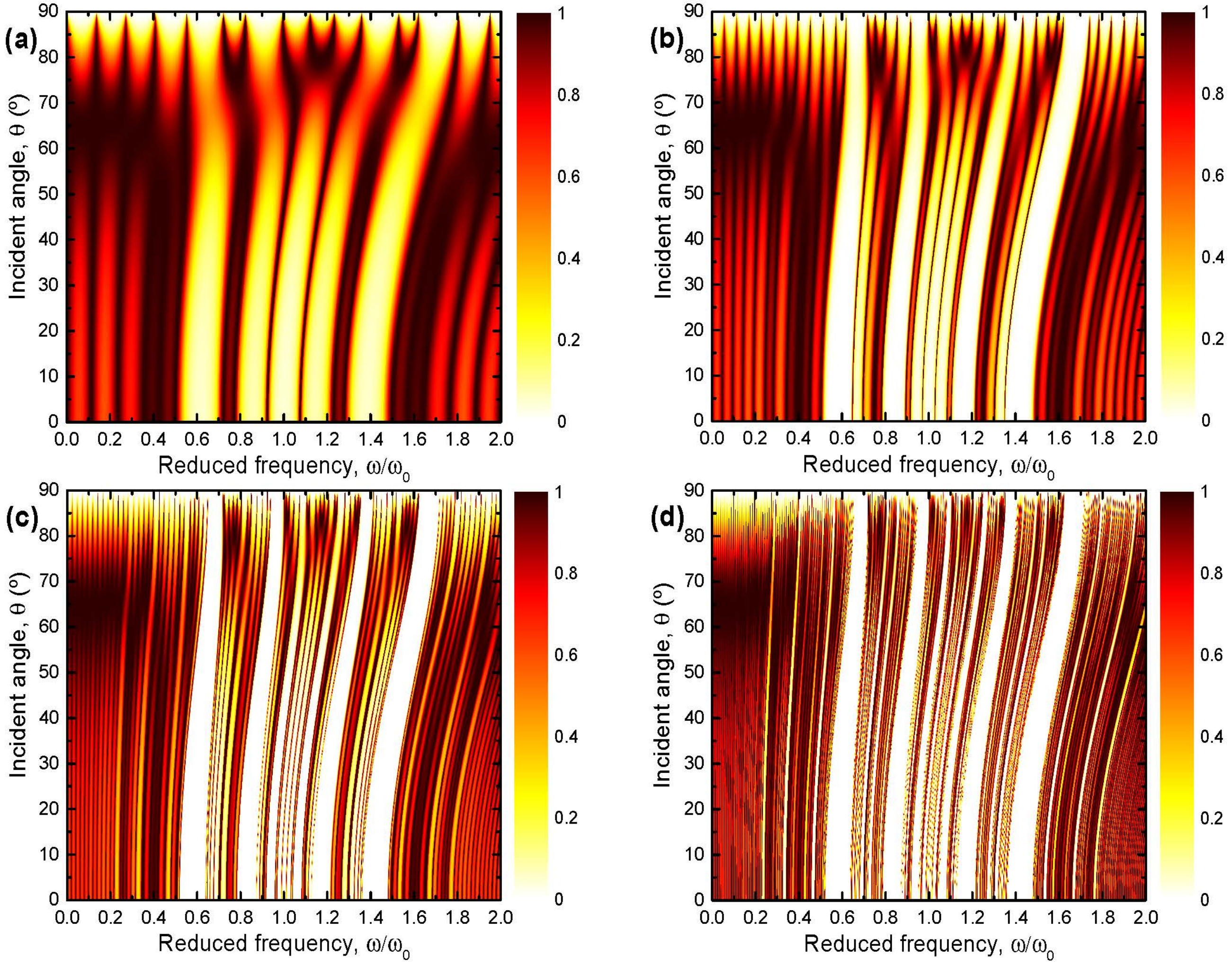}
 \caption{Same as Fig. \ref{fig3}, but now for TM waves ($p$-polarization). For this case, a well defined omnidirectional photonic band gap does not appear.}\label{fig4}
\end{figure}

In Fig. \ref{fig3}, we have the transmission coefficient versus the reduced frequency $\omega/\omega_{0}$ and the incident angle $\theta_{C}$ for 4th, 5th, 6th and 7th generations (Figs. \ref{fig3}a, \ref{fig3}b, \ref{fig3}c and \ref{fig3}d), respectively, for TE polarization. On the other hand, in Fig. \ref{fig4} we have the same as Fig. \ref{fig3}, but for TM polarization. Only for TE waves we could observe a narrow region with a well defined photonic band gap, enhanced by the green rectangle. Beside of this, the band gaps are independent of the incident angle, and are all localized at, approximately, the same reduced frequency region ($0.6\le \omega/\omega_{0}\le 0.65$) and they have almost the same width. Clearly, the appearing of these band gaps are due to the long-range order of the arrangements of the layers in the Octonacci lattice, and this long-range order is responsible for the anomalous interference, giving rise to these omnidirectional band gaps.

\section{Conclusions}\label{sec:Conclusions}

In summary, in this work we calculated the transmission spectra for photonic crystals generated in accordance to Octonacci sequence. Firstly, for normally incident wave, we observe that, for a same generation, the transmission spectra for both TE and TM waves are equal, at least qualitatively, and they present a scaling property where a self-similar behavior is obtained as a qualitative evidence of these spectra are fractals. Our results also show the spectra regions where the omnidirectional band gaps emerges for specific generations of Octonacci photonic structures, except to TM waves. For TE waves, we note that all of them have the almost same width, for different generations. Also, the localization of modes is enhanced, as we increase the generation number, as it was expected\cite{Eudenilson2}. This property could serve as a tool for sensing devices, since the localization can be controlled through physical parameters as the thickness of the layers\cite{Barth}. In this case, it is expected that the quasiperiodicity could affect in someway the transmission spectra, creating well defined band gaps and a self-similar pattern as in Fibonacci multilayered photonic structure\cite{Vasconcelos}.

There are many applications of such type of systems in optical
system and communications. For example, recently several authors
have studied one dimensional quasicrystals in order to understood
the passband \cite{Xu}, absorption \cite{Liu}, and Omnidirectional
reflection \cite{Barriuso} properties in Fibonacci (generalized in
last case) quasicrystals. Besides the selfsimilarity, they have
found that is possible to control the bandgap properties by
constructing the photonic crystal heterostructure with photonic
crystals and quasi-crystals \cite{Xu}. By using the effective medium
theory,  Xu et al \cite{Xu} have concluded that is possible to
understand of the passband of the photonic crystal and
quasi-crystal, and the optimal lengths for the maximum bandgap and
phase matching condition can be obtained. On other hand, the study
of the absorption properties in one dimensional quasiperiodic
photonic cystals composed by a thin film metallic layer and
dielectric Fibonacci multilayer, have resulted to almost perfect
absorption for the energy of the incident wave, by tuning the length
of the dielectric layer together with the high refractive index
\cite{Liu}, making it possible to design a perfect absorber with
channel number as high as twelve and bandwidth as narrow as 1 nm, which is very attractive in sensor/detector technology and
highly integrated dense wavelength division multiplexing networks.
Also, the omnidirectional reflection is possible in generalized
Fibonacci quasicristals by controlling the thickness \cite{Barriuso}.
All these ideas could be extended to photonic crystals generated in accordance to the Octonacci sequence
studied here. We hope with our work to stimulate others researchers to
investigate experimentally and theoretically this interesting
quasiperiodic system in other to foster further applications.

\begin{acknowledgments}
The authors would like to thank CAPES and CNPq  (Brazilian Science Funding Agencies) for the financial support.
\end{acknowledgments}


\begin{thebibliography}{99}

\bibitem{Yablonovitch} J.E. Yablonovitch, ``Inhibited Spontaneous Emission in Solid-State Physics and Electronics," Phys. Rev. Lett. \textbf{58}, 2059-2062 (1987).

\bibitem{John} S. John,``Strong localization of photons in certain disordered dielectric superlattices," Phys. Rev. Lett. \textbf{58}, 2486-2489 (1987).

\bibitem{Prather}  D.W. Prather, A. Sharkawy, S. Shi, J. Murakowski, and G. Schneider, \textit{Photonic Crystals: Theory, Applications and Fabrication}, Willey, (2009).

\bibitem{Laine} V.E. Laine (ed.), \textit{Photonic Crystals: Fabrication, Band Structure and Applications}, Nova Science Pub. Inc., (2010).

\bibitem{Massaro} A. Massaro (ed.), \textit{Photonic Crystals - Introduction, Applications and Theory}, InTech, (2012).

\bibitem{Sakoda} K. Sakoda, \textit{Optical Properties of Photonic Crystals}, 2nd ed., Springer, (2004).

\bibitem{Joannopoulos} J.D. Joannopoulos, S.G. Johnson, J.N. Winn, and R.D. Meade, \textit{Photonic Crystals: Molding the Flow of Light}, 2nd ed., Princeton University Press, (2008).

\bibitem{Cerqueira} A. Cerqueira Jr.,``Recent progress and novel applications of photonic crystal fibers," Rep. Prog. Phys. \textbf{73}, 024401 (2010).

\bibitem{Macia} E. Maci\'a,``Exploiting aperiodic designs in nanophotonic devices," Rep. Prog. Phys. \textbf{75}, 036502 (2012).

\bibitem{Kramper} P. Kramper, M. Agio, C.M. Soukoulis, A. Birner, F. Muller, R.B. Wehrspohn, U. Gosele, and V. Sandoghdar, ``Highly Directional Emission from Photonic Crystal Waveguides of Subwavelength Width," Phys. Rev. Lett. \textbf{92}, 113903 (2004).

\bibitem{Chigrin} D.N. Chigrin, A.V. Lavrinenko, D.A. Yarotsky, and S.V. Gaponenko, ``Observation of total omnidirectional reflection from a one-dimensional dielectric lattice," Appl. Phys. A: Mater. Sci. Process. \textbf{68}, 25-28 (1999).

\bibitem{Shechtman} D. Shechtman, I. Blech, D. Gratias, and J. W. Cahn.,``Metallic Phase with Long-Range Orientational Order and No Translational Symmetry," Phys. Rev. Lett. \textbf{53}, 1951 (1984).

\bibitem{Eudenilson2} E.L. Albuquerque, and M.G. Cottam, \textit{Polaritons in Periodic and Quasiperiodic Structures}, Elsevier, Amsterdam, (2004).

\bibitem{Shechtman2} Page of 2011 Nobel Prize in Chemistry awarded to Dan Shechtman: \verb"http://www.nobelprize.org/nobel_prizes/chemistry/"\newline \verb"laureates/2011/press.html".

\bibitem{Bohr} H. Bohr,``Fastperiodische Funktionen," Jahresbericht der Deutschen Mathematiker-Vereinigung \textbf{34},
25-40 (1926).

\bibitem{Steurer} W. Steurer, and S. Deloud, \textit{Crystallography of Quasicrystals -- Concepts, Methods and Structures}, Springer, (2009).

\bibitem{Penrose} R. Penrose, ``The Role of Aesthetics in Pure and Appied Mathematical Research," Bulletin of the Institute of Mathematics and its Applications \textbf{10},
266-271 (1974).

\bibitem{Grunbaum} B. Gr\"{u}nbaum, and G.C. Shephard, \textit{Tilings and Patterns}. W.H. Freeman, New York, (1986).

\bibitem{Ammann} R. Ammann, B. Gr{\"{u}}nbaum, and G.C. Shephard, ``Aperiodic Tiles," Discrete \& Computational Geometry \textbf{8}, 1-25 (1992).

\bibitem{Kramer} P. Kramer and R. Neri,``On periodic and non-periodic space fillings of E$^m$ obtained by projection," Acta Crystallographica Section A \textbf{40}, 580-587 (1984).

\bibitem{Towle} R. Towle. \textit{Colored Zonotiles}, from Wolfram Library Archive, http://library.wolfram.
com/infocenter/MathSource/1197/

\bibitem{Macia2} E. Maci\'a,``The role of aperiodic order in science and technology," Rep. Prog. Phys. \textbf{69}, 397-441 (2006).

\bibitem{Steurer2} W. Steurer, and D.S.-Widmer,``Photonic and phononic quasicrystals," J. Phys. D: Appl. Phys. \textbf{40}, R229-R247 (2007).

\bibitem{Vardeny} Z.V. Vardeny, A. Nahata, and A. Agrawal,``Optics of photonic quasicrystals," Nat. Photonics \textbf{7}, 177-187 (2013).

\bibitem{Costa} C.H.O. Costa, P.H.R. Barbosa, F.F. Barbosa Filho, M.S. Vasconcelos, and E.L. Albuquerque, ``Band gaps in the terahertz frequency range in quasiperiodic one-dimensional magnonic crystals," Solid State Commun. \textbf{150}, 2325-2328 (2010).

\bibitem{Kohmoto} M. Kohmoto, B. Sutherland, and K. Iguchi,``Localization of optics: Quasiperiodic media," Phys. Rev. Lett. \textbf{58}, 2436-2438 (1987).

\bibitem{Zentgraf} T. Zentgraf, A. Christ, J. Kuhl, N.A. Gippius, S.G. Tikhodeev, D. Nau, and H. Giessen, ``Metallodielectric photonic crystal superlattices: Influence of periodic defects on transmission properties," Phys. Rev. B \textbf{73}, 115103 (2006).

\bibitem{Utikal} T. Utikal, T. Zentgraf, S.G. Tikhodeev, M. Lippitz, and H. Giessen, ``Tailoring the photonic band splitting in metallodielectric photonic crystal superlattices," Phys. Rev. B \textbf{84}, 075101 (2011).

\bibitem{Barber_book_2008} E.M. Barber (ed.), \textit{Aperiodic Structures in Condensed Matter: Fundamentals and Applications}, CRC Press, (2008).

\bibitem{dal_Negro_book_2013} L. dal Negro (ed.), \textit{Optics of Aperiodic Structures: Fundamentals and Device Applications}, Pan Stanford Publishing, (2013).

\bibitem{Costa3} C.H.O. Costa, and M.S. Vasconcelos, ``Band gaps and transmission spectra in generalized Fibonacci $\sigma(p,q)$ one-dimensional magnonic quasicrystals," J. Phys.: Condens. Matter \textbf{25}, 286002 (2013).

\bibitem{Costa4} C.H. Costa, and M.S. Vasconcelos, ``Magnons in one-dimensional k-component Fibonacci structures," J. Appl. Phys. \textbf{115}, 17C115 (2014).

\bibitem{Vasconcelos} M.S. Vasconcelos, E.L. Albuquerque, and A.M. Mariz, ``Optical localization in quasi-periodic multilayers," J. Phys.: Condens. Mat. \textbf{10}, 5839-5849 (1998).

\bibitem{Vasconcelos3} M.S. Vasconcelos, and E.L. Albuquerque, ``Transmission fingerprints in quasiperiodic dielectric multilayers," Phys. Rev. B \textbf{59}, 11128-11131 (1999).

\bibitem{prb62} H.Q. Yuan, U. Grimm, P. Repetowicz, and M. Schreiber, ``Energy spectra, wave functions, and quantum diffusion for quasiperiodic systems," Phys. Rev. B \textbf{62}, 15569-15578 (2000).

\bibitem{Lu} J.P. Lu, T. Odagaki, and J.L. Birman, ``Properties of one-dimensional quasilattices," Phys. Rev. B \textbf{33}, 4809-4817 (1986).

\bibitem{Barriuso} A. G. Barriuso, J. J. Munz\'{o}n, T. Yonte, A. Felipe, and L. L. S\'anchez-Soto, ``Omnidirectional reflection from generalized Fibonacci quasicrystals," Opt. Express \textbf{21},
30039 (2013).

\bibitem{Golmohammadi} S. Golmohammadi, M.K. Moravvej-Farshi, A. Rostami, and A. Zarifkar, ``Spectral analysis of fibonacci-class one-dimensional quasi-periodic structures," Progress In Electromagnetics Research \textbf{75}, 69-84 (2007).

\bibitem{Zhan} T. Zhan, X. Shi, Y. Dai, X. Liu, and J. Zi, ``Transfer matrix method for optics in graphene layers," J. Phys.: Condens. Matter \textbf{25}, 215301 (2013).

\bibitem{Gellermann} W. Gellermann, M. Kohmoto, B. Sutherland, and P. C. Taylor, ``Localization of light waves in Fibonacci dielectric multilayers," Phys. Rev. Lett. {\bf 72}, 633-636 (1994).

\bibitem{Barth} S. Barth, F.H.-Ramirez, J.D. Holmes, and A.R.-Rodriguez, ``Synthesis and applications of one-dimensional semiconductors," Prog. Mater. Sci. \textbf{55}, 563-627 (2010).

\bibitem{Xu} S. Xu, Y. Zhu, L. Wang, P. Yang, and P.K. Chu, ``Passband and defective bands in photonic and quasi-crystals," J. Opt. Soc. Am. B \textbf{31},
664-671(2014).

\bibitem{Liu} Y. Gong, X. Liu, L. Wang, H. Lu, and G. Wang, ``Multiple responses of TPP-assisted near-perfect absorption in metal/Fibonacci quasiperiodic photonic crystal," Optics Express \textbf{19}, 9759-9769 (2011).

\end{thebibliography}
\end{document}